\documentclass[twoside,fleqn]{ActaStyle}
\usepackage{times}

\usepackage{lscape}
\usepackage{graphicx,epsfig}
\usepackage[tbtags]{amsmath}

\def\authorheading{U.C.~T\"auber}

\def\shorttitle{Dynamic phase transitions in diffusion-limited reactions}

\begin{document}

\pagerange{1}{9}

\title{DYNAMIC PHASE TRANSITIONS IN DIFFUSION-LIMITED REACTIONS}

\author{Uwe C. T\"auber\email{tauber@vt.edu}}
{Department of Physics, Virginia Tech, Blacksburg, VA 24061-0435, USA}

\day{May 15, 2002}

\abstract{
Many non-equilibrium systems display dynamic phase transitions from 
active to absorbing states, where fluctuations cease entirely. 
Based on a field theory representation of the master equation, the 
critical behavior can be analyzed by means of the renormalization 
group.
The resulting universality classes for single-species systems are
reviewed here.
Generically, the critical exponents are those of directed percolation
(Reggeon field theory), with critical dimension $d_c = 4$. 
Yet local particle number parity conservation in even-offspring
branching and annihilating random walks implies an inactive phase
(emerging below $d_c' \approx 4/3$) that is characterized 
by the power laws of the pair annihilation reaction, and leads to 
different critical exponents at the transition.
For local processes without memory, the pair contact process with 
diffusion represents the only other non-trivial universality class.
The consistent treatment of restricted site occupations and quenched 
random reaction rates are important open issues.}


\section{Introduction: Active to absorbing state phase transitions}
\label{sec:intr} \setcounter{section}{1}\setcounter{equation}{0}

Among the prevalent goals in current statistical mechanics is the 
understanding and characterization of non-equilibrium steady states.
This program is hindered by the general absence of an effective 
free-energy function that would allow a straightforward classification 
in terms of symmetries and interactions. 
One might hope, however, that the task becomes more feasible near
continuous phase transitions separating different non-equilibrium 
steady states. 
For in analogy with equilibrium critical points, one would expect the 
properties near non-equilibrium phase transitions as well to be 
{\em universal}, i.e., independent of the detailed microscopic 
dynamical rules and the initial conditions.
Rather, the emerging power laws and scaling functions describing the 
long-wavelength, long-time limit should hopefully be characterized by 
not too many distinct universality classes.
Obviously, provided we can cast the problem at hand into a form 
amenable to field-theoretic methods, the renormalization group (RG)
provides a very powerful tool for such investigations.
Indeed, during the past twenty-five years or so it has been 
successfully applied to a variety of non-equilibrium processes.
Among the lessons we have learned is that critical phenomena or 
generic scale invariance far from thermal equilibrium, where the
detailed-balance constraints do not apply, are considerably richer than 
equilibrium statics, or even near-equilibrium dynamics.
In fact, intuitions drawn from the latter may often be quite deceptive.

A special class of genuine non-equilibrium phase transitions separate 
{\em `active'} from {\em `inactive, absorbing'} stationary states where 
any stochastic fluctuations cease entirely \cite{chopard98,hinrichsen00}.
These occur in a large variety of systems, e.g., in chemical reactions 
involving an inert state $\emptyset$ that does not release the 
reactants $A$ anymore.
We may also consider stochastic population dynamics, combining, say, 
diffusive migration with asexual reproduction $A \to 2 A$ (with rate 
$\sigma$), spontaneous death $A \to \emptyset$ (rate $\mu$), and 
lethal competition $2 A \to A$ (rate $\lambda$). 
In the inactive state, where no population members $A$ are left, all 
processes terminate.
Similar effective dynamics may be used to model non-equilibrium 
physical systems, such as the domain-wall kinetics in Ising chains
with competing Glauber (spin flip) and Kawasaki (spin exchange) 
dynamics \cite{grassberger84}. 
Here, spin flips $\uparrow \uparrow \downarrow \downarrow \, \to \, 
\uparrow \uparrow \uparrow \downarrow$ and $\uparrow \uparrow 
\downarrow \uparrow \, \to \, \uparrow \uparrow \uparrow \uparrow$ may 
be viewed as domain wall ($A$) hopping and pair annihilation $2 A \to 
\emptyset$, respectively, whereas a spin exchange $\uparrow \uparrow 
\downarrow \downarrow \, \to \, \uparrow \downarrow \uparrow 
\downarrow$ represents a branching process $A \to 3 A$ in domain wall 
language. 
Notice that the paramagnetic and ferromagnetically ordered phases 
map onto the active and inactive `particle' states, the latter 
rendered absorbing if the spin flip rates are computed at zero 
temperature, allowing no energy increase.

The simplest mathematical description for such processes uses a 
kinetic rate equation in terms of the time-dependent average 
`particle' density $n(t)$, which for the above systems reads
\begin{equation}
  \partial_t \, n(t) = \left( \sigma - \mu \right) n(t) - \lambda \, 
  n(t)^2 \ . \label{mfreq}
\end{equation}
Obviously, this yields both an inactive and an active phase, as for 
$\sigma < \mu$ we have $n(t \to \infty) \to 0$, whereas 
for $\sigma > \mu$ the particle density saturates at $n_s = (\sigma 
- \mu) / \lambda$.
The explicit solution $n(t) = n_0 n_s / \left[ n_0 + (n_s - n_0) 
e^{(\mu - \sigma) t} \right]$ shows that either stationary state is 
approached exponentially in time.
The two phases are separated by a continuous dynamic transition at 
the critical point $\sigma = \mu$, where the temporal decay becomes
algebraic, $n(t) = n_0 / (1 + n_0 \lambda t)$.
However, eq.~(\ref{mfreq}) entails a mean-field type of approximation, 
as we have neglected particle correlations on the right-hand side, 
and effectively factored a two-point correlation function, namely the
joint probability of finding two particles at the same position.
A more detailed treatment therefore requires a systematic 
incorporation of spatio-temporal fluctuations, and the ensuing 
particle correlations.

\section{Langevin description, Reggeon field theory, and 
	 directed percolation (DP)}
\label{sec:dper}

We may try a phenomenological incorporation of fluctuation effects
using a Langevin-type approach.
To this end, we assume diffusive particle transport, and model our 
`chemical' system through a non-linear stochastic differential 
equation with reaction functional $r[n]$,
\begin{equation}
  \partial_t \, n({\bf x},t) = D \, \nabla^2 \, n({\bf x},t) - 
  r[n]({\bf x},t) + \zeta({\bf x},t) \ . \label{recfn}
\end{equation}
The stochastic variable $\zeta({\bf x},t)$ incorporates the reaction 
noise, which we take to have zero mean, and to be local in space-time 
(`white' in Fourier space), with a density-dependent correlator $c[n]$:
\begin{equation}
  \langle \zeta \rangle = 0 \ , \quad 
  \langle \zeta({\bf x},t) \, \zeta({\bf x}',t') \rangle = c[n] \, 
  \delta({\bf x}-{\bf x}') \, \delta(t-t') \ . \label{noifn}
\end{equation}
Here, the last expression means that whenever a trajectory average is
taken, $c[n]$ needs to be factored in.
In the spirit of Landau theory, we may expand the functionals $r[n]$ 
and $c[n]$ near the inactive phase ($n \ll 1$).
In the absence of spontaneous particle production, both must vanish
at $n=0$, which is the condition for an absorbing state.
Consequently, an active to absorbing state phase transition should
be {\em generically} described by eqs.~(\ref{recfn}), (\ref{noifn}) 
with $r[n] \approx D r \, n + \lambda \, n^2$ and $c[n] \approx 2 
\sigma \, n$ \cite{grassberger78}.
Upon neglecting fluctuations, one then recovers eq.~(\ref{mfreq}) with 
$r = (\mu - \sigma)/D$.

By means of standard techniques, a stochastic differential equation 
of the form (\ref{recfn}) with noise correlation (\ref{noifn}) can 
be represented as a functional integral \cite{janssen76}.
Essentially, one starts with the Gaussian noise probability 
distribution $W[\zeta] \propto \exp \left[ - \int (\zeta^2 / 2 c) 
d^dx \, dt \right]$, multiplies with $1 = \int D[n] \prod_{{\bf x},t} 
\delta(\partial_t n - D \nabla^2 n + r - \zeta) = \int D[i {\tilde n}] 
D[n] \exp \left[ - \int {\tilde n} (\partial_t n - D \nabla^2 n + r - 
\zeta) d^dx \, dt \right]$, thus introducing the auxiliary fields 
${\tilde n}$, and then integrates out the stochastic noise $\zeta$.
Upon applying a forward discretization, the ensuing functional
determinant vanishes, and one arrives at the probability distribution
$P[n] \propto \int D[i {\tilde n}] \exp (- S[{\tilde n},n])$, with 
the response functional $S[{\tilde n},n] = \int {\tilde n} \left( 
\partial_t n - D \nabla^2 n + r[n] - c[n] {\tilde n}/2 \right) d^dx \, 
dt$. After rescaling $n = (\sigma/\lambda)^{1/2} \phi$, ${\tilde n} = 
(\lambda/\sigma)^{1/2} {\tilde \phi}$, the expanded Langevin equation 
(\ref{recfn}), (\ref{noifn}) maps to {\em `Reggeon' field theory} 
\cite{migdal74}
\begin{equation}
  S[{\tilde \phi},\phi] = \int \left[ {\tilde \phi} \left[ \partial_t 
  + D ( r - \nabla^2) \right] \phi + u \left( {\tilde \phi} \phi^2 - 
  {\tilde \phi}^2 \phi \right) \right] d^dx \, dt \ , \label{rftac}
\end{equation}
with $u = (\sigma \lambda/2)^{1/2}$. 
Notice the invariance of this action with respect to `rapidity 
inversion' $\phi({\bf x},t) \to - {\tilde \phi}({\bf x},-t)$, 
${\tilde \phi}({\bf x},t) \to - \phi({\bf x},-t)$.

\begin{table}[b]
\begin{center}
\begin{tabular}{|c||c|c|c|}
\hline
DP exponents & $d=1$ & $d=2$ & $d=4-\epsilon$, $O(\epsilon)$ \\ \hline
$n_s \sim |r|^\beta$         & $\beta \approx 0.2765$  
& $\beta \approx 0.584$      & $\beta = 1-\epsilon/6$     \\
$\xi \sim |r|^{-\nu}$        & $\nu \approx 1.100$ 
& $\nu \approx 0.735$        & $\nu = 1/2+\epsilon/16$    \\
$t_c \sim \xi^z \sim |r|^{-z \nu}$ & $z \approx 1.576$ 
& $z \approx 1.73$           & $z = 2-\epsilon/12$        \\ 
$n_c(t) \sim t^{-\alpha}$    & $\alpha \approx 0.160$
& $\alpha \approx 0.46$      & $\alpha = 1-\epsilon/4$    \\ \hline 
\end{tabular} 
\caption{Critical exponents for the saturation density (order 
  parameter) $n_s$, correlation length $\xi$, characteristic time 
  scale $t_c$, and critical density decay $n_c(t)$ for the 
  universality class of directed percolation (DP).} \label{dpcexp}
\end{center}
\end{table}
The field theory (\ref{rftac}) should capture the generic critical 
behavior for non-equilibrium phase transitions between active and 
absorbing states, occurring at $r = 0$.
Quite remarkably, the very same action is obtained for the threshold 
pair correlation function \cite{cardy80} in the purely geometric 
problem of {\em directed percolation} (DP) \cite{kinzel83}.
Power counting reveals $d_c = 4$ as the upper critical dimension;
hence, the critical exponents as predicted by mean-field theory 
acquire logarithmic corrections at $d_c$, and are shifted to different 
values by the infrared-singular fluctuations in $d < 4$ dimensions.
By means of the standard perturbational loop expansion in terms of the
diffusion propagator and the vertices $\propto u$, and the application 
of the RG, the critical exponents can be computed systematically and 
in a controlled manner in a dimensional expansion with respect to 
$\epsilon = 4 - d$.
The one-loop results, to first order in $\epsilon$, as well as 
reliable values from Monte Carlo simulations in one and two dimensions
\cite{hinrichsen00} are listed in Tab.~\ref{dpcexp}.
Moreover, as a consequence of rapidity invariance there are only three 
independent scaling exponents, namely the anomalous field dimension 
$\eta$, the correlation length exponent $\nu$, and the dynamic critical 
exponent $z$.
All other exponents are then fixed by scaling relations, such as 
$\beta = \nu (d + z - 2 + \eta) / 2 = z \nu \alpha$ for the order 
parameter exponent $\beta$ and the critical density decay exponent 
$\alpha$, respectively.

\section{From the master equation to stochastic field theory}
\label{sec:meft}

The above phenomenological approach constitutes a natural extension
of the mean-field rate equations to a stochastic partial differential
equation.
However, it does presume (i) that such a Langevin-type representation 
is in fact possible, and (ii) it is fundamentally based on conjectures
on the noise correlator.
Away from thermal equilibrium, there is no analog of the Einstein 
relation to constrain the form and structure of the noise correlations.
Indeed, it has emerged that the latter may quite profoundly affect the 
scaling behavior of non-equilibrium systems, rendering point (ii) an
entirely non-trivial issue.
Moreover, as we shall see, even assertion (i) turns out to be relevant
in certain important model systems.
It is therefore of fundamental importance to be able to construct a
long-wavelength or field theory representation of stochastic processes
that starts directly from their microscopic definition in terms of a
classical master equation, without recourse to any serious additional 
assumptions or approximations.

Fortunately, for reaction-diffusion systems there exists indeed a 
standard route from the master equation to an effective `Hamiltonian'
(more precisely, the Liouville time evolution operator), and therefrom 
immediately to a field theory action \cite{doi76}.
The key point is that all possible configurations here can be
labeled by specifying the occupation numbers $n_i$ of, say, the sites
of a $d$-dimensional lattice; we shall henceforth assume that there
are no site occupation restrictions, i.e., $n_i = 0, 1, 2, \ldots$
The master equation then addresses the time evolution of the 
configurational probability $P(\{ n_i \};t)$.
For example, the corresponding contribution from the binary 
coagulation process $2 A \to A$ at site $i$ reads $\partial_t P(n_i;t) 
|_\lambda = \lambda \left[ (n_i+1) n_i P(n_i+1;t) - n_i (n_i-1) 
P(n_i;t) \right]$. 
This sole dependence on the integer variables $\{ n_i \}$ calls for a
second-quantized bosonic operator representation with the standard 
commutation relations $[ a_i , a_j^\dagger ] = \delta_{ij}$ and the 
empty state $| 0 \rangle$ such that $a_i | 0 \rangle  = 0$.
We then define the Fock states via $| \{ n_i \} \rangle = \prod_i 
(a_i^\dagger)^{n_i} | 0 \rangle $ (notice that the normalization is
different from standard many-particle quantum mechanics), and thence
construct the formal state vector $| \Phi(t) \rangle = 
\sum_{\{ n_i \}} P(\{ n_i \};t) \, | \{ n_i \} \rangle$.
The master equation now imposes a linear time evolution that can be
written as an imaginary-time `Schr\"odinger' equation 
$\partial_t | \Phi(t) \rangle = - H \, | \Phi(t) \rangle$, with a 
generally non-Hermitian stochastic `Hamiltonian' 
$H(\{a_i^\dagger\},\{a_i\})$; e.g., for the on-site coagulation 
reaction one finds explicitly $H_{\lambda \, i} = - \lambda (1 - 
a_i^\dagger) a_i^\dagger a_i^2$.

Our goal is to evaluate time-dependent statistical averages for 
observables $F$, necessarily mere functions of the occupation 
numbers as well, whence $\langle F(t) \rangle = \sum_{\{ n_i \}} 
F(\{ n_i \}) P(\{ n_i \};t)$.
Straightforward algebra utilizing the identity $[e^a,a^\dagger] = e^a$
shows that this average can be cast into a `matrix element' 
$\langle F(t) \rangle = \langle {\cal P} | F(\{ a_i \}) | \Phi(t) 
\rangle  = \langle {\cal P} | F(\{ a_i \}) \, e^{- H t}| \Phi(0) 
\rangle$ with the state vector $| \Phi(t) \rangle$ and the projector 
state $\langle {\cal P} | = \langle 0 | \prod_i e^{a_i}$, with
$\langle {\cal P} | 0 \rangle = 1$.
For example, probability conservation implies $1 = \langle {\cal P} | 
e^{- H t}| \Phi(0) \rangle$, i.e., for infinitesimal times $\langle 
{\cal P} | \Phi(0) \rangle = 1$ and $\langle {\cal P} | H = 0$, which 
is satisfied if $H(\{1\},\{a_i\}) = 0$.
Notice furthermore that commuting the factor $e^{\sum_i a_i}$ through
all the other operators has the effect of shifting $a_i^\dagger \to
1 + a_i^\dagger$ everywhere.

As a final step, we employ the coherent-state path integrals familiar 
from quantum many-particle systems \cite{popov81}, and perform the 
continuum limit to arrive at the desired field theory.
For our earlier population dynamics example with diffusive motion
and the reactions $A \to 2 A$, $A \to \emptyset$, and $2 A \to A$,
this results in the action (omitting contributions from the initial
state)
\begin{equation}
  S[{\hat \psi},\psi] = \int \left[ {\hat \psi} \left( \partial_t - 
  D \, \nabla^2 \right) \psi + \sigma (1 - {\hat \psi}) {\hat \psi} 
  \psi - \mu (1 - {\hat \psi}) \psi - \lambda (1 - {\hat \psi}) 
  {\hat \psi} \psi^2 \right] d^dx \, dt \ . \label{dpmft}
\end{equation}
It is important to emphasize again that the single approximation used
here was the continuum limit; specifically, no assumptions whatsoever
on the stochastic noise were invoked.
The mean-field rate equation (\ref{mfreq}) is recovered by inserting 
the solution $\hat \psi = 1$ to the stationarity condition 
$\delta S / \delta \psi = 0$ into $\delta S / \delta {\hat \psi} = 0$,
with $n(t) = 2 \langle \psi({\bf x},t) \rangle$.
However, higher moments of the field $\psi$ cannot be directly 
identified with the corresponding density correlations.
Very remarkably, though, after shifting ${\hat\psi}(x,t) = 1 + 
{\tilde \psi}(x,t)$, such that $\langle {\tilde \psi} \rangle = 0$, 
and appropriate field rescaling one arrives again at the field theory 
(\ref{rftac}) with $r = (\mu - \sigma)/D$ and $u = (2 \sigma
\lambda)^{1/2}$, with the additional vertex $\lambda \, {\tilde \phi}^2
\phi^2$.
Yet, the coupling $\lambda$ has scaling dimension $\kappa^{2-d}$,
where $\kappa$ denotes a momentum scale, whereas $u$ becomes marginal
at $d_c = 4$.
Thus, at least in the vicinity of the upper critical dimension, 
$\lambda$ constitutes an irrelevant variable in the RG sense, and in 
general the ratio $\lambda / u \sim \kappa^{-d/2}$ is expected to 
scale to zero asymptotically.
Consequently, for processes in the directed percolation universality
class the microscopic master equation and the coarse-grained Langevin 
description both lead to the effective action of Reggeon field theory.

\section{Diffusion-limited annihilation}
\label{sec:dlan}

Let us now investigate the $k$-th order {\em annihilation} reaction 
$k A \to \emptyset$ \cite{lee94}.
The corresponding mean-field rate equation reads $\partial_t \, n(t) = 
- \lambda \, n(t)^k$.
For simple radioactive decay ($k = 1$), it is solved by the familiar 
exponential $n(t) = n_0 \, e^{-\lambda_1 t}$, whereas one obtains 
power laws for $k \geq 2$, namely $n(t) = \left[ n_0^{1-k} + (k-1) 
\lambda \, t \right]^{-1/(k-1)}$.
In order to consistently include fluctuations in the latter case, we
start out from the master equation once again, which leads to the 
action
\begin{equation}
  S[{\hat \psi},\psi] = \int \left[ {\hat \psi} \left( \partial_t - 
  D \nabla^2 \right) \psi - \lambda (1 - {\hat \psi}^k) \psi^k \right] 
  d^dx \, dt \ . \label{anmft}
\end{equation}
After shifting ${\hat\psi}(x,t) = 1 + {\tilde \psi}(x,t)$, it becomes
clear that this field theory actually has no simple Langevin 
representation.
For in order to interpret ${\tilde \psi}$ as the corresponding noise
auxiliary field, it should appear quadratically in the action only,
and with negative prefactor.
Thus, the Langevin equation derived from the action (\ref{anmft})
for the pair annihilation process would imply unphysical `imaginary' 
noise with $c[n] = - 2 \lambda n^2$.

Analyzing the field theory (\ref{anmft}) further, we see that the 
diffusion propagator does not become renormalized at all, implying
that $\eta = 0$ and $z = 2$ to all orders in the perturbation 
expansion.
The critical dimension of the annihilation vertex is found to be
$d_c(k) = 2/(k-1)$, allowing for the possibility of non-trivial 
scaling behavior in low physical dimensions only for the pair and 
triplet processes.
The simple structure of the action permits summing the entire 
perturbation series for the vertex renormalization by means of a 
Bethe-Salpeter equation that in Fourier space reduces to a geometric 
series \cite{lee94}.
For pair annihilation ($k = 2$), this yields the following 
asymptotic behavior for the particle density: $n(t) \propto t^{-1}$ 
for $d > 2$, $n(t) \propto t^{-1} \ln t$ at $d_c = 2$, and $n(t) 
\propto t^{-d/2}$ for $d < 2$.
The slower decay for $d \leq 2$ originates in the fast mutual
annihilation of any close-by reactants; after some time has elapsed, 
this leaves only well-separated particles.
The annihilation dynamics thus produces {\em anti}-correlations, 
mimicking effective repulsion (which is also the physical 
interpretation of the negative sign in $c[n]$).
Similarly, for triplet annihilation ($k = 3$) the density decays as
$n(t) \propto t^{-1/2}$ for $d > 1$, with mere logarithmic 
corrections $n(t) \propto \left( t^{-1} \ln t \right)^{1/2}$ at
$d_c = 1$.

\section{Branching and annihilating random walks (BARW)}
\label{sec:barw}

In order to allow again for a genuine phase transition, we combine 
the annihilation $k A \to \emptyset$ ($k \geq 2$) with {\em branching} 
processes $A \to (m+1) \, A$.
The associated rate equation reads $\partial_t \, n(t) = - \lambda 
\, n(t)^k + \sigma \, n(t)$, with the solution $n(t) = n_s / \left( 
1 + \left[ (n_s / n_0)^{k-1} - 1 \right] e^{- (k-1) \sigma t} 
\right)^{1/(k-1)}$.
Mean-field theory thus predicts the density to approach the saturation 
value $n_s = (\sigma / \lambda)^{1/(k-1)}$ as $t \to \infty$ for any
positive branching rate.
Above the critical dimension $d_c(k) = 2/(k-1)$ therefore, the system 
only has an {\em active} phase; $\sigma_c = 0$ represents a degenerate 
`critical' point, with scaling exponents essentially determined by
the pure annihilation model: $\alpha = \beta = 1/(k-1)$, $\nu = 1/2$, 
and $z = 2$.
However, Monte Carlo simulations for these {\em branching and
annihilating random walks} (BARW) revealed a much richer picture, in 
low dimensions clearly distinguishing between the cases of {\em odd} 
and {\em even} number of offspring $m$ \cite{grassberger84,takayasu92}:
For $k = 2$, $d \leq 2$, and $m$ {\em odd}, a transition to an 
inactive, absorbing phase is found, characterized by the DP critical 
exponents.
On the other hand, for {\em even} offspring number a phase transition
in a novel universality class with $\alpha \approx 0.27$, $\beta 
\approx 0.92$, $\nu \approx 1.6$, and $z \approx 1.75$ emerges in one 
dimension.

The aforementioned mapping to a stochastic field theory, combined with
RG methods, succeeded to elucidate the physics behind those remarkable 
findings \cite{cardy96}.
The action for the most interesting pair annihilation case becomes 
\begin{equation}
  S[{\tilde \psi},\psi] = \int \left[ {\hat \psi} \left( \partial_t - 
  D \, \nabla^2 \right) \psi - \lambda \, (1 - {\hat \psi}^2) \, \psi^2 
  + \sigma \, (1 - {\hat \psi}^m) \, {\hat \psi} \psi \right] d^dx \, dt 
  \ . \label{bawft}
\end{equation}
Upon combining the reactions $A \to (m+1) A$ and $2 A \to \emptyset$,
one notices immediately that the loop diagrams generate the lower-order 
branching processes $A \to (m-1) A$, $A \to (m-3) A \ldots$
Moreover, the one-loop RG eigenvalue $y_\sigma = 2 - m(m+1)/2$ 
(computed at the annihilation fixed point) shows that the reactions 
with smallest $m$ are the most relevant.
For {\em odd} $m$, we see the generic situation is given by $m = 1$, 
i.e., $A \to 2 A$, supplemented with the spontaneous decay $A \to 0$.
After a first coarse-graining step, this latter process (with rate 
$\mu$) must be included in the effective model, which hence becomes 
identical with action (\ref{dpmft}).
We are thus led to Reggeon field theory (\ref{rftac}) describing the
DP universality class, {\em provided} the induced decay processes are
sufficiently strong to render $\sigma_c > 0$.
Yet for $d > 2$ the renormalized mass term $\sigma_R - \mu_R$ remains
positive, which leaves us with merely the active phase captured by 
mean-field theory.
For $d \leq 2$, however, the involved fluctuation integrals are 
infrared-divergent, thus indeed allowing the induced decays to 
overcome the branching processes to produce a phase transition.
As the DP upper critical dimension is $d_c = 4$, the scaling exponents 
display an unusual discontinuity at $d = 2$, jumping from the 
non-trivial two-dimensional DP to the mean-field values (for any
$d > 2$) as a consequence of the vanishing critical branching rate 
\cite{cardy96}.

It is now obvious why the case of {\em even} offspring number $m$ is 
fundamentally different: 
Here, the most relevant branching process is $A \to 3 A$, and 
spontaneous particle death with associated exponential decay is 
{\em not} generated under the coarse-grained dynamics, which precludes 
the above mechanism for producing an inactive phase with exponential 
particle decay.
This important distinction from the odd-$m$ case can be traced to a 
microscopic local conservation law, for the reactions $2 A \to 
\emptyset$ and $A \to 3 A$, $A \to 5 A \ldots$ always destroy or 
produce an even number of reactants, and consequently preserve the 
particle number parity.
Formally, this is reflected in the invariance of the action 
(\ref{bawft}) under the combined inversions $\psi \to -\psi$,
${\hat \psi} \to -{\hat \psi}$.
As we saw earlier, the branching rate $\sigma$ certainly constitutes a
relevant variable near the critical dimension $d_c = 2$.
Therefore the phase transition can only occur at $\sigma_c = 0$, and
for any $\sigma > 0$ there exists only an active phase, described by
mean-field theory.
In two dimensions one readily predicts the following logarithmic
corrections: $\xi(\sigma) \propto \sigma^{-1/2} \ln (1/\sigma)$, and
$n(\sigma) \propto \sigma \, [\ln (1/\sigma)]^{-2}$.  
However, setting $m = 2$ in the one-loop value for the RG eigenvalue 
$y_\sigma$, we notice that the branching vertex becomes irrelevant
for $d < 4/3$.
More information on the low-dimensional behavior can be gained through
a one-loop analysis at {\em fixed} dimension, albeit uncontrolled 
\cite{cardy96}.
The ensuing RG flow equations for the renormalized, dimensionless 
branching rate $\sigma_R = \sigma / D \kappa^2$, and annihilation rate
$\lambda_R = C_d \lambda / D \kappa^{2-d}$, with $C_d = \Gamma(2-d/2)
/ 2^{d-1} \pi^{d/2}$ read (for $m = 2$):
\begin{equation}
  \frac{d \sigma_R}{d \ell} = \sigma_R \left[ 2 - 
  \frac{3 \, \lambda_R}{(1 + \sigma_R)^{2-d/2}} \right] \ , \quad 
  \frac{d \lambda_R}{d \ell} = \lambda_R \left[ 2 - d - 
  \frac{\lambda_R}{(1 + \sigma_R)^{2-d/2}} \right] \ . \label{bawfeq}
\end{equation}
The effective coupling is then identified as $g = \lambda_R / (1 + 
\sigma_R)^{2-d/2}$, which approaches the annihilation fixed point
$g^* = 2 - d$ as $\sigma_R \to 0$, while for $\sigma_R \to \infty$
the flow tends towards the Gaussian fixed point $g^* = 0$ describing
the active state.
The separatrix between the two phases is given by the {\em unstable}
RG fixed point $g^* = 4 / (10 - 3d)$, which enters the physical regime
below the novel critical dimension $d_c' \approx 4/3$.
For $d < d_c'$, this describes a dynamic phase transition with
$\sigma_c > 0$; the emerging {\em inactive} phase is characterized by
a vanishing branching rate, and thus by the {\em algebraic} scaling
behavior of the pure pair annihilation model, $n(t) \propto t^{-d/2}$.
This very different character of the inactive phase also explains why
the transition cannot possibly fall into the DP category.
The aforementioned fixed-dimension RG analysis yields the rather crude
values $\nu \approx 3 / (10 - 3 d)$, $z \approx 2$, and 
$\beta \approx 4 / (10 - 3 d)$ for this {\em parity-conserving} (PC)
universality class. 
Moreover, the absence of any mean-field ounterpart for this transition
precludes a sound derivation of `hyperscaling' relations such as 
$\beta = z \nu \alpha$.

Similar arguments suggest for triplet annihilation $3 A \to \emptyset$ 
combined with branching $A \to (m+1) A$ that DP behavior with 
$\sigma_c > 0$ should ensue for $m \ {\rm mod} \ 3 = 1,2$, as then the 
processes $A \to \emptyset$, $A \to 2 A$, and $2 A \to A$ are 
dynamically generated.
For $m = 3, 6, \ldots$, on the other hand, because of $d_c(k=3) = 1$
one expects logarithmic corrections only in one-dimensional systems.

\section{Concluding remarks and outlook}
\label{cncl}

In this brief overview, I have outlined how non-linear stochastic 
processes can be represented by field theory actions, allowing for a
thorough analysis and classification by means of the renormalization 
group.
Systems with a single `particle' species displaying a non-equilibrium
phase transition from an active to an inactive, absorbing state are 
generically captured by the DP universality class.
The second prominent example, applicable when additional symmetries
(degenerate absorbing states) are present, is that of branching and
annihilating random walks with even offspring number (PC).
In fact, there appears to be only one additional possibility for
non-trivial scaling behavior in `bosonic' systems with multiple site
occupation, namely the combination of {\em binary} reactions $2 A \to 
\emptyset$ and $2 A \to (m+2) A$ (annihilation and fission; pair 
contact process with diffusion, PCPD) \cite{grassberger82}.
These reactions subsequently generate $2 A \to m A, (m-2) A , \ldots$, 
$2 (m+1) A , \ldots$, thus producing {\em infinitely} many couplings 
with identical scaling dimensions, which renders the field theory 
non-renormalizable.
Yet, the master equation immediately leads to the exact evolution
equation $\partial_t \, n(t) = (\sigma - \lambda) \langle m (m-1) 
\rangle$, which establishes the existence of a phase transition when 
the particle production and annihilation rates precisely balance.
In the inactive, absorbing phase, obviously the power laws of the pure
annihilation ($m$ even) or coagulation ($m$ odd) models are recovered.
In the active phase, the particle density diverges in a finite time,
and the transition itself is {\em discontinuous}.
These irregular features are `cured' by restricting the site 
occupation numbers, whereupon the phase transition becomes continuous,
with critical exponents $\alpha \approx 0.13$, $\beta \approx 0.58$, 
$z \approx 1.83$ that belong to neither the DP nor PC universality
classes.
Yet different exponents were recently found for the triplet reactions
$3 A \to 2 A$ and $3 A \to 4 A$ \cite{park02}.
It is however unclear at present how to generally and systematically 
incorporate such site restrictions into a field theory description in 
a tractable manner \cite{brunel00}.
Another open problem concerns the obviously very relevant issue of 
quenched disorder in the reaction rates.
For example, the RG investigation for DP with random percolation
threshold yields run-away flows to diverging couplings, with an as
yet unclear interpretation \cite{moreira96}.

Lastly, the rich variety of multi-component systems is only partially 
understood.
Only DP processes with arbitrarily many particle species have been
fully classified, namely any non-linear coupling leads directly to 
the DP universality class again \cite{janssen97}, whereas novel 
multi-critical behavior may ensue for hierarchical systems comprising
spontaneous transformations $A \to B$, $B \to C \ldots$ 
\cite{tauber98,janssen97}.
Yet already in the simple binary process $A + B \to \emptyset$, the
initial conditions play a vital role.
For equal $A$ and $B$ densities, the fact that $n_A - n_B$ remains 
conserved under the dynamics leads to segregation and reaction-front 
dominated kinetics, with shifted upper critical dimension $d_c = 4$ 
and asymptotic decay $n_A \propto n_B \propto t^{-d/4}$ for lower
dimensions \cite{lee95}.
On the other hand, whenever there is a majority species, the minority
decays to zero exponentially, although this may be masked by very long 
crossover times.
The (bosonic) $N$-species BARW generalization $A_i + A_i \to \emptyset$,
$A_i \to A_i + 2 A_j$ turns out to be exactly analyzable, with a 
degenerate phase transition only at vanishing branching rate, and 
critical exponents $\alpha = d/2$, $\beta = 1$, $\nu = 1/d$, 
$z = 2$ \cite{cardy96}.
The situation for $N=1$ is thus qualitatively different from all
multi-component cases.
Moreover, in certain cases the difference in diffusion constants is a 
relevant control parameter \cite{wijland98}.
When there are `passive' agents, the corresponding variables are 
readily integrated out, leaving however interactions of the remaining 
degrees of freedom that are non-local in time.
A classic example for such memory effects is dynamic percolation 
\cite{janssen85}, which at criticality yields stationary isotropic
percolation clusters. 
A recent argument claims this to govern the generic universality class 
for non-equilibrium phase transitions with infinitely many absorbing 
states \cite{wijland02}.
Many intriguing problems are still open, and RG techniques will be 
invaluable tools for the further analysis of scale-invariant 
non-equilibrium systems.

\begin{ack}
I thank the organizers of RG02 for inviting me to this inspiring meeting
in the beautiful High Tatra Mountains.
I am very grateful to John Cardy for introducing me to the topic 
discussed here, and gladly acknowledge fruitful and pleasant 
collaborations and discussions with him, as well as with Olivier 
Deloubri\'ere, Yadin Goldschmidt, Geoff Grinstein, Malte Henkel, 
Henk Hilhorst, Haye Hinrichsen, Martin Howard, Hannes Janssen, 
G\'eza \'Odor, Klaus Oerding, Zolt\'an R\'acz, Beate Schmittmann, 
Gunter Sch\"utz, Steffen Trimper, Fred van Wijland, and Royce Zia.       
Over the years, this research has been supported by the Engineering and 
Physical Sciences Research Council (GR/J78327), the European Commission 
(ERB FMBI-CT96-1189), the Deutsche Forschungsgemeinschaft (Ta 177/2), 
the National Science Foundation (DMR 0075725), and the Jeffress Memorial 
Trust (J-594).
\end{ack}

\end{document}